\documentstyle{article}
\begin{document}
\large
\baselineskip=24pt
\begin{center}
{\Large
\bf A Lattice Boltzmann Model for Multi-phase Fluid Flows}
\end{center}

\vspace{0.3in}

\begin{center}
Daryl Grunau$^{1,3}$, Shiyi Chen$^{2}$, and Kenneth Eggert$^{1}$
\end{center}

\vspace{0.3in}

${}^{1}${\footnotesize Earth and Environment Science Division,
Los Alamos National Laboratory, Los Alamos, NM 87545}

${}^{1}${\footnotesize Theoretical Division and Center for
Nonlinear Studies, Los Alamos, NM 87545}

${}^{3}${\footnotesize Department of Mathematics,
Colorado State University, Fort Collins, CO 80523}

\vspace{0.4in}
\begin{center}
{\bf ABSTRACT}
\end{center}

We develop a lattice Boltzmann equation method for simulating multi-phase
immiscible fluid flows with variation of density and viscosity,
based on the model proposed
by Gunstensen {\em et al}\cite{gun} for two-component immiscible fluids.
The numerical measurements of surface tension and viscosity
agree well with theoretical predictions.  Several basic numerical tests,
including spinodal decomposition, two-phase fluid flows in two-dimensional
channels and two-phase viscous fingering,
are shown in agreement of experiments and analytical solutions.

\pagebreak
\section{Introduction}
It has recently been shown that lattice gas computational fluid dynamics,
including lattice gas automata (LGA) \cite{fhp}
and lattice Boltzmann (LB)\cite{zanetti} equation methods,
provide alternative numerical techniques for solving
the Navier-Stokes equations, multi-phase fluid flows\cite{roth1,somers,syc0}
and a variety of other fluid systems\cite{gary1,syc1}.
The parallel nature of these newly developed schemes, adapted from
cellular automata, affords an easy implementation of
fast, efficient and accurate simulations on parallel machines.

To solve the incompressible fluid equations by traditional numerical methods
such as finite differences or finite elements, one must deal
with a Poisson equation for the pressure term that is induced by the continuum
condition and the momentum equation.  This fact has brought to light a
crucial difference between lattice gas methods and traditional algorithms,
since a lattice gas solver is only required to solve the kinetic equation
of the particle distribution.  All other quantities,
including density, velocity and energy can be obtained by a
macroscopic averaging process through the
distribution function.  Therefore, the pressure effects
on the momentum equation are controlled by an equation of state.

Strictly speaking, lattice gas automata should be restricted to solving
compressible fluid flows.  The incompressible limit theory of
lattice gas methods\cite{fhp1} and their numerical simulations\cite{syc1},
however, predict and demonstrate that such a lattice gas
solution at low Mach number produces results comparable with those of
the incompressible Navier-Stokes
equations for a wide variety of problems and with very small numerical
errors.  Also, two-phase surface dynamic boundaries and wall boundaries
are far more easily implemented with a lattice gas method
compared to direct simulations of the incompressible Navier-Stokes
equations\cite{roth1,syc1}.

Rothman and Keller\cite{roth1} were the first to extend the single-phase
lattice gas model proposed by Frisch, Hasslacher and Pomeau\cite{fhp}
to simulate multi-phase fluid problems.  Colored particles were
introduced to distinguish between phases, and a nearest-neighbor particle
interaction was used to facilitate interfacial dynamics, such as Laplace's
formula for surface tension.  Later, Somers and Rem\cite{somers},
and Chen {\em et al}\cite{syc0} extended the original colored particle
scheme by introducing colored holes.  It has been shown\cite{syc0}
that the colored-hole lattice gas method
extends the original nearest neighbor particle interaction
to several lattice lengths, leading to a Ukowa potential.  Moreover,
the colored-hole scheme carries purely local information in it's particle
collision step, reducing the size of the look-up table in the algorithm
and consequently speeding up the simulation.

Although two-phase lattice gas algorithms are able to produce interesting
surface phenomena, they are difficult to compare quantitatively
with experiments and other numerical simulations due to their noisy
characteristics induced by particle fluctuations. The lattice Boltzmann
model proposed by McNamara and Zanetti\cite{zanetti}, however, solves the
kinetic equation for the particle {\em distribution} instead of tracking
each particle's motion, as is done in lattice gas methods.  Nonetheless,
lattice gas algorithms will be superior to lattice Boltzmann models
for simulating underlying physics and modeling microscopic dynamics such as
correlation effects and phase transitions, since they contain more information
about the microscopic behavior of particles.  On the other hand, to provide
a numerical method for solving partial differential equations governing
macroscopic behavior, lattice Boltzmann schemes will be at least
as good as lattice gas models.  The finite-difference nature of
lattice Boltzmann methods not only can simulate the macroscopic equations
more efficiently and accurately, but also preserve some of the advantages
of lattice gas models, such as their parallel computing nature and their
ease of boundary implementation.

Combining the lattice Boltzmann model of McNamara and Zanetti\cite{zanetti}
and the two-phase lattice gas model of Rothman {\em et al}, Gunstensen
{\em et al} \cite{gun} proposed a lattice Boltzmann method for solving
two-phase fluid flows. An important contribution of this model is the
introduction of a perturbation step (shown in detail below) so that
Laplace's formula at an interface can be approximately recoved.  This
lattice Boltzmann method has been used in several applications \cite{gun1},
however it has a few fundamental problems.  First, the model does not solve
the exact two-phase fluid equations -- although Galilean invariance is
recovered by the proper assignment of rest particles, the equation of
state remains to be velocity dependent\cite{syc2}.
Secondly, the model uses a fully linearized collision operator, which involves
a 24*24 matrix multiplication at each time step and position in three
dimensional space, reducing computational efficiency.  Finally, the model is
restricted to two-phase fluid flows having the same densities and viscosities.
In this paper, we extend their model by using a single-time relaxation
lattice Boltzmann model.  With the proper choice of the particle equilibrium
distribution function, we are able to recover the incompressible Navier-Stokes
equations for the color-blind fluid.  In addition, our new model makes
provision for variable densities and viscosities by use of the freedom of
the rest particle equilibrium distribution, and a space dependent
relaxation process.  For simplicity we present a two-phase,
two-dimensional fluid model on a hexagonal lattice in this paper.
The extension of the current model to multi-phase fluids and other
lattices, including a two-dimensional square lattice, a three-dimensional
Face-Centered Hypercubic lattice\cite{fhp1} and a Body-Centered-Cubic
\cite{syc3} lattice, will be presented in detail in another
paper\cite{daryl}.

\section{Numerical Model}

Denote $f_i({\bf x},t), f_i^{(r)}({\bf x},t)$ and $f_i^{(b)}$ as the particle
distribution functions at space ${\bf x}$ and time $t$ for total, red,
and blue fluids respectively.  Here $i = 0, 1,\cdot\cdot\cdot,N$,
where $N$ is the number of moving particle directions (6 for the hexagonal
lattice and 8 for the square lattice in two dimensions), and
$f_i = f_i^{(r)} + f_i^{(b)}$.  The lattice Boltzmann equation for both
red and blue fluids can be written as follows:
\begin{equation}
f_i^{k}({\bf x} + {\bf e}_i , t +1) = f_i^{k}({\bf x}, t) +
\Omega_i^{k}({\bf x}, t), \label{eq:1}
\end{equation}
were $k$ denotes either the red or blue fluid, and $\Omega_i^{k} =
(\Omega_i^{k})^{1} + (\Omega_i^{k})^{2}$ is the collision operator.
Note that in the two-phase lattice Boltzmann model by Gunstensen
{\em et al} only a color-blind kinetic equation was given.
The first term of the collision operator, $(\Omega_i^{k})^{1}$,
represents the process of
relaxation to local equilibrium.  For simplicity, we use a linearized
collision operator with a single time relaxation parameter $\tau_k$\cite{syc4},
\[ (\Omega_i^{k})^{1} = \frac{1}{\tau_k}(f_i^{k}-f_i^{k(eq)}). \]
Here $f_i^{k(eq)}$ is the local equilibrium
state depending on the local density and velocity, and $\tau_k$ is
the characteristic relaxation time for species $k$.  In lattice gas automaton
models, the form of the local equilibrium state and rate of relaxation to
the local equilibrium distribution
are completely determined by the particle scattering process.
 From a statistical point of view, the macroscopic effects of
these collisions only arise in the transport coefficients.
Therefore, lattice Boltzmann models ignore particle collision details
such as collision cross sections and frequencies.  In addition, the
local equilibrium state can be arbitrarily chosen\cite{syc2}, with the
exception
that it must satisfy the conservation of mass and momentum:
\[ \rho_r = \sum_i f_i^{r} = \sum_i f_i^{r(eq)}, \]
\[ \rho_b = \sum_i f_i^{b} = \sum_i f_i^{b(eq)}, \]
and
\[ \rho {\bf v} = \sum_{i, k} f_i^{k} {\bf e}_i =
\sum_{i, k} f_i^{k(eq)} {\bf e}_i. \]
Here $\rho_r\ {\rm and}\ \rho_b$ are densities of the red and blue fluids
respectively, $\rho =
\rho_r + \rho_b$ is the total density and ${\bf v}$ is the local velocity.

Following the pressure-corrected Galilean-invariant single-phase
lattice Boltzmann model proposed by reference \cite{syc2}, we assume
the following equilibrium distribution for both red and blue fluids:
\begin{eqnarray}
f_i^{r(eq)}&=&\rho_r/(6 + m_r) + \rho_r({\bf e}_i \cdot {\bf v}/3 +
2/3({\bf e}_i)_\alpha({\bf e}_i)_\beta v_\alpha v_\beta-
\frac{1}{6}{\bf v}^2), \nonumber\\
f_0^{r(eq)}&=&\rho_r m_r/(6 + m_r)- \rho_r {\bf v}^2, \nonumber \\
f_i^{b(eq)}&=&\rho_b/(6 + m_b) + \rho_b({\bf e}_i \cdot {\bf v}/3 +
2/3({\bf e}_i)_\alpha({\bf e}_i)_\beta v_\alpha v_\beta-
\frac{1}{6}{\bf v}^2), \nonumber\\
f_0^{b(eq)}&=&\rho_b m_b/(6 + m_b)- \rho_b {\bf v}^2. \label{eq:2}
\end{eqnarray}
Here we have introduced two parameters, $m_r$ and $m_b$.  They can be
understood as the number of degenerate rest particle states for the
red and blue fluids respectively.  To achieve a stable interface,
we furthermore assume that the moving particle distributions for both red
and blue fluids are the same when ${\bf v} = 0$:, i.e., $d=\rho_r/(6+m_r)
=\rho_b/(6+m_b)$.  This implies the following density ratio: $ \frac{\rho_r}
{\rho_b} = \frac{6 + m_r}{6 + m_b}$.

The second part of the collision operator is similar to
that given in Gunstensen {\em et al}'s paper:
\[ (\Omega_i^{k})^{2} = \frac{A_k}{2}|{\bf F}| (({\bf e}_i \cdot {\bf F})^2
/|{\bf F}|^2 - 1/2), \]
where ${\bf F}$ is the local color gradient, defined as:
\[ {\bf F}({\bf x}) = \sum_i {\bf e}_i (\rho_r({\bf x}+{\bf e}_i) -
\rho_b({\bf x}+{\bf e}_i)). \]
Note that in a single-phase region of our incompressible fluid model,
${\bf F} = 0$.
Thus the second term of the collision operator, $(\Omega_i^{k})^{2}$,
only has contribution at two-phase interfaces.  The parameter $A_k$ is
a free parameter, controlling the surface tension (shown before).
For simplicity, we choose $A_r = A_b$ in this paper.

To obtain a surface tension, we follow Rothman's scheme\cite{roth1} to
redistribute colored particles at two-phase interfaces
(without changing the total particle distribution, $f_i$) by enforcing
the red color momentum ${\bf j}^{r} = \sum_{i} f_i^{r} {\bf e}_i$,
to align with the direction of the
local color gradient.  In other words, we redistribute the red density at
an interface to maximize the following quantity:
\[ -({\bf j}^{r} \cdot {\bf F}). \]
The blue particle distribution can then be recovered using:
$f_i^{b} = f_i - f_i^{r}$.

To derive hydrodynamics, we use a long-wavelength, low-frequency
approximation and a multiscaling analysis as follows:
\[ \frac{\partial}{\partial t} =
\epsilon \frac{\partial}{\partial t_{1}}+
\epsilon^{2} \frac{\partial}{\partial t_{2}} + ..., \]
\[ \frac{\partial}{\partial x} = \epsilon \frac{\partial}{\partial x_1}. \]
Here $t_1$ and $t_2$ represent fast and slow time scales respectively, and
$\epsilon$ is assumed to be a small expansion parameter.

A Taylor series expansion of Equation (\ref{eq:1})
to second order in the lattice spacing and time step gives the following
continuum kinetic equation:
\begin{eqnarray}
\frac{\partial f_i^{k}}{\partial t}
+{\bf e}_i \cdot {\bf \nabla}f_i^{k}
+\frac{1}{2} {\bf e}_{i} {\bf e}_{i} :{\bf \nabla}
{\bf \nabla} f_i^{k}
+{\bf e}_{i}
\cdot {\bf \nabla}\frac{\partial}{\partial t}f_i^{k}
+\frac{1}{2}\frac{\partial}{\partial t}\frac{\partial}{\partial t}
f_i^{k} = \Omega_{i}^{k}.
\end{eqnarray}

Taking the zero and first order moments of ${\bf e}_i$ over equation
(\ref{eq:2}), we readily obtain the continuum equations for the red
and blue fluids, and the following momentum equation for the color-blind
(total) fluid:
\[ \frac{\partial (\rho v_{\alpha})}{\partial t} +
\sum_{i,k} {\bf e}_{\alpha}{\bf e}_{i} (f_i^{k(eq)} +
f_i^{k(1)}) = 0, \]
where $f_i^{k(1)}$ is the next order perturbation for $f_i^{k}$.
Note that since $$\sum_i (\Omega_i^{k})^{2} = \sum (\Omega_i^{k})^{2}
{\bf e}_i= 0$$ and $${\bf F} = \frac{1}{3} \nabla (\rho_r - \rho_b) +
\cdot\cdot\cdot \sim \epsilon + o(\epsilon^2),$$
the second term in the collision operator does not
contribute to the continuum and momentum equations of
the first order approximation.  It, however, will contribute to
the pressure term at an interface:
$P = P_0 + \epsilon |\nabla F|$ , where $P_0$ comes from the equation
of state.

Our two-phase immiscible fluid model contains three fluid regions: a red
or blue homogeneous region and a thin region where the two fluids mix
at the interfaces. In a homogeneous region, the evolution
of our model will recover
the Navier-Stokes equations with viscosity of $\nu_k = (2\tau_k- 1)/8$
and sound speed of $c_k = \sqrt{3/(6+m_k)}$ according to the
Chapman-Enskog expansion shown in reference \cite{syc3}.  A variation
of viscosities for the two fluids can be obtained by choosing
different $\tau_k$.

The last issue remaining to be addressed is the interfacial dynamics that
take place in a region where red and blue fluids are adjacent.
Usually the thickness of such an interface will depend on an averaged
relaxation time, $2\tau_r \tau_b / (\tau_r + \tau_b)$, and the rest
particle distribution.  There are
several ways to construct a relaxation parameter so that the red and
blue fluids have a smooth change of viscosity at their interfaces.
To do this, we define an order parameter, $\psi$, depending on
red and blue densities as follows: $\psi=\frac{\rho_r - \rho_b}{\rho_r +
\rho_b}$.  Note that in general $|\psi| \le 1 $, however in a purely red
fluid region, $\psi = 1$, and in a purely blue fluid region, $\psi = -1$.
To continuously connect relaxation parameters $\tau_r$ and $\tau_b$
at an interface, we employ the following simple formula:
\[ \tau = \left\{ \begin{array}{ll} \tau_r & \mbox{ $\psi > \delta$}\\
g_r(\psi) & \mbox{$\delta \ge \psi > 0$} \\
g_b(\psi) & \mbox{$0 \ge \psi \ge -\delta$} \\
\tau_b & \mbox{$\psi < -\delta$} \end{array} \right. \]
where $g_r(\psi)$ and $g_b(\psi)$ are second order functions of $\psi$:
\[ g_r(\psi) = \alpha + \beta \psi + \gamma \psi^2, \] and
\[ g_b(\psi) = \alpha + \eta \psi + \xi \psi^2. \]
Assuming that $g_r(\delta) = \tau_r, g_b(-\delta) = \tau_b$, $\frac{\partial
\tau}{\partial \psi} = 0$ at $|\psi| = \delta$ and
that $g_r(0) = g_b(0)$, we have: $\alpha = 2
\tau_r \tau_b / (\tau_r + \tau_b), \beta = 2 (\alpha - \tau_b)/\delta
, \gamma = \beta/(2 \delta), \eta = 2(\tau_r - \alpha)/\delta$ and
$\xi = \eta / (2\delta)$.  Here $\delta$ is a free parameter controlling
the interface thickness, $\delta \le 1$.

Using the standard definition of surface tension\cite{gun},
$\sigma = \int (P_n - P_t)dx$, and after some algebra\cite{daryl}, we obtain
a theoretical formula for the surface tension on a two-dimensional
hexagonal lattice:  $\sigma=9A<\tau> d (12 + m_r + m_b),$ where $<\tau>$
is the averaged
relaxation time step across the interface.  Here the integration is over whole
space along the direction perpendicular to a given interface, and $P_n$ and
$P_t$ are respectively the normal and tangential stress tensor at the two-phase
interface.  For the interface relaxation interpolation scheme in this paper,
we use  $<\tau> = 2\tau_r \tau_b/(\tau_r + \tau_b)$.  In Fig. 1, we show
the theoretical prediction (---) and numerical measurements
($\diamond$ and +) of $\sigma$ as a function of particle density $d$ with
a mass ratio, $m_r/m_b = 2$.  The $\diamond$ symbols represent numerical
measurements of surface tension using the above
definition, and the + signs represent the surface tension obtained by
Laplace's formula with measurements of the pressure drop across a
bubble.  As seen, the theoretical prediction and numerical measurements
agree very well in both cases.

\section{Numerical Simulations and Discussion}

To demonstrate the application of the presented lattice Boltzmann model,
we show three numerical examples in the following sections.  These simulations
have been carried out
using the CM-200 at the Advanced Computing Laboratory at Los Alamos National
Laboratory.  The first numerical test is the study of the phase
segregation process by two-dimensional spinodal decomposition when the fluids
have differing densities.  Fig. 2 is a snapshot ($t=8300$) of the
two-phase area distribution when the density ratio, $\gamma=
\frac{m_r}{m_b},{\rm\ is}\ 10$.  One can see that the
red fluid represents a high density region and the blue fluid represents
a low density region.  The initial condition for this simulation is
constant density and random color distribution with periodic boundaries.
Here the surface tension parameter is set to $A=0.01$.
The current model preserves the basic two-phase segregation processes
and, in simulations not shown here, we have seen the preservation of
stable two-phase interfaces for much higher density ratios ($\sim 200$).
This capability will allow us to approach the simulation of gas and oil
mixing flows.

The second numerical test is two-dimensional channel flow with two fluids
having different viscosities.  A lattice length of 65 in the y-direction
($W=65$) and 128 in the x-direction ($L=128$) is used, with a non-slip
condition on both the upper and bottom boundaries.  Initially, the lower
half space (from lattice 1 to 32) is filled with red fluid, and the top half
space (from lattice 34 to 65) is filled with blue fluid.
The middle line is initialized to half red and half blue, and
the relaxation parameters for red and blue fluids are assigned values
of $\tau_r = 2$ and $\tau_b = 1$.
Using the viscosity formula discussed above, this translates into a
kinematic viscosity ratio of 3:1 for the red and blue fluids
respectively.  A forcing technique, as described in reference \cite{syc3},
is used to establish a small pressure gradient across the length of
the channel.  Assuming a thin interface in the middle, an analytical
solution of the velocity distribution as a function of $y$ can be
derived, centered on the middle of the channel\cite{tran}:
\begin{eqnarray}
v_x^{r}&=&\frac{(P_0-P_L)w^2}{2\mu_1L}
\left [ \frac{2\mu_r}{\mu_r + \mu_b} +
\frac{\mu_r-\mu_b}{\mu_r+\mu_b}\frac{y}{w}-\frac{y}{w}^2\right ], \nonumber \\
v_x^{b}&=&\frac{(P_0-P_L)w^2}{2\mu_2L}\left [ \frac{2\mu_b}{\mu_r+\mu_b} +
\frac{\mu_r-\mu_b}{\mu_r+\mu_b}\frac{y}{w}-\frac{y}{w}^2\right ],
\end{eqnarray}
where $(P_0-P_L)/L$ is the pressure gradient across the
channel, $\mu_r$ and $\mu_b$ are shear viscosities for red (0.7875) and
blue (0.2625) fluids respectively.  Here also $w=\sqrt{3}W/4$
is half the channel width, where the factor $\sqrt{3}/2$ is a hexagonal
lattice effect.  In Fig. 3 we present the analytical prediction (---)
and direct numerical simulation ($\diamond$) for velocity as a function
of channel width $y$.  As seen, the agreement is excellent.

The last numerical test is a two dimensional Hele-Shaw viscous fingering
experiment. The upper and lower walls are assigned a no-slip boundary
condition, and the fluids are given a viscosity ratio of 1:10 by
assigning $\tau_r=1$ and $\tau_b=5.5$.  To develop a Hele-Shaw pattern,
an initial perturbation is introduced at the interface\cite{perturb,degregoria}
which is then forced by maintaining a high pressure at the inlet and low
pressure at the outlet.  The shape of the finger can be altered by a
change in the surface tension parameter $A$, given as $0.0065$ in this
experiment.  Fig. 4 shows the evolution of a less viscous fluid penetrating
one having a higher viscosity at times = 0, 2000, 4000, 6000, 8000 and 10000.
We see that a stable finger develops and is maintained throughout the
simulation, although we expect a tip-splitting instability to occur for
smaller values of $A$, as demonstrated by reference \cite{degregoria}.
This phenomenon will be discussed in more detail in a subsequent paper.
The fluid patterns shown here are similar to results obtained by other
numerical methods\cite{perturb,degregoria}.

The lattice Boltzmann model in this paper is the extension of the model
proposed
by Gunstensen {\em et al}.  The current model gives an exact Navier-Stokes
solution, in the incompressible limit, for each fluid individually.
The surface tension at an interface satisfies the Laplace formula, and
the model has the capability to simulate two-phase fluid flows with different
densities and viscosities.  The simulation results for several typical
two-phase fluid flows are shown, in good agreement with analytical solutions
and some experimental observations.  Several issues, however, remain to be
addressed.  First, to achieve a density variation, the current model
uses the freedom of the rest particle equilibrium distribution.  Although the
scheme is simple, it is difficult to obtain high velocity flows since
the assignment of a high particle mass ratio will accumulate a large
number of particles in the rest state, reducing the fluid speed.  In addition,
the equation of state of the current model is that of an ideal gas
similar to the original FHP model.  To simulate fluid flows with a
large density variation, a lattice gas model with an equation of state
for non-ideal gases (liquids) may
provide some alternatives.  Second, the current model uses two parabolic
curves to match the interface relaxation characteristic time in the mixing
region.  For a very thin interface (for example, one lattice width),
the parameter, $\delta$, will be negligible.  On the other hand, for flows
with slow relaxation, the interface will be much wider.
The relaxation form and the choice of $\delta$ may affect interfacial
dynamics.  Other functional forms, such as a bi-normal distribution,
may give better results.

We thank G. D. Doolen and Hudong Chen for
useful suggestions.  Eugene Loh has contributed to the original
CM-2 code development.  This work is supported by the US Department of Energy
at Los Alamos National Laboratory.  Numerical simulation were carried out
using the computational resources at the
Advanced Computing Laboratory at the Los Alamos National Laboratory.

\section{Figure Captions}
\begin{description}

\item[Fig. 1]  The theoretical prediction (---) and numerical measurements
($\diamond$ and +) of surface tension $\sigma$ as a function of particle
density $d$ with a mass ratio $\gamma = 2$.

\item[Fig. 2] A snapshot at time $t=8300$ of the area density distribution
of a spinodal decomposition simulation where the two fluids have a 10:1
mass ratio, $\gamma$.   The boundaries are periodic, and the initial
condition is assigned a constant density with a random color
distribution.  The surface tension parameter is assigned to $A=0.01$.

\item[Fig. 3]The analytical prediction (---) and direct numerical
simulation ($\diamond$) of velocity as a function of channel width $y$.

\item[Fig. 4]The formation of a stable finger in a Hele-Shaw cell at times
$t =$ 0, 2000, 4000, 6000, 8000 and 10000.  A viscosity ratio of
1:10 is established by setting $\tau_r=1$ and $\tau_b=5.5$.  The
low-viscosity fluid (red) is penetrating the higher viscosity region (blue).

\end{description}

\end{document}